\documentclass[english,aps,prl,twocolumn,nofootinbib,preprintnumbers,groupedaddress,10pt]{revtex4-1}

\usepackage[latin1]{inputenc}
\usepackage{graphicx}
\usepackage{amsmath,amsthm,amssymb}
\usepackage{enumitem}

\usepackage{dcolumn}
\usepackage{hyperref}
\usepackage{bm}

\def\be{\begin{equation}}
\def\ee{\end{equation}}
\def\ba{\begin{eqnarray}}
\def\ea{\end{eqnarray}}
\def\bs{\begin{subequations}}
\def\es{\end{subequations}}

\usepackage{color}

\begin{document}
\preprint{IPMU12-0118}

\title{Massive gravity: nonlinear instability of the homogeneous and isotropic universe}

\author{Antonio De Felice}
\affiliation{ThEP's CRL, NEP, The Institute for Fundamental Study, Naresuan University, Phitsanulok 65000, Thailand}
\affiliation{Thailand Center of Excellence in Physics, Ministry of Education, Bangkok 10400, Thailand}

\author{A. Emir G\"umr\"uk\c{c}\"uo\u{g}lu}
\author{Shinji Mukohyama}
\affiliation{Kavli Institute for the Physics and Mathematics of the Universe, Todai Institutes for Advanced Study, University of Tokyo, 5-1-5 Kashiwanoha, Kashiwa, Chiba 277-8583, Japan}

\begin{abstract}
 We argue that all homogeneous and isotropic solutions in nonlinear
 massive gravity are unstable. For this purpose, we study the
 propagating modes on a Bianchi type--I manifold. We analyze their
 kinetic terms and dispersion relations as the background manifold
 approaches the homogeneous and isotropic limit. We show that in this
 limit, at least one ghost always exists and that its frequency tends to
 vanish for large scales, meaning that it cannot be integrated out from
 the low energy effective theory. This ghost mode is interpreted as a 
 leading nonlinear perturbation around a homogeneous and isotropic
 background. 
\end{abstract}

\date{\today}

\maketitle

{\bf Introduction.~} 
The concept of the mass has been central in many areas of
physics. Gravitation is not an exception, and it is one of the simplest
but yet unanswered questions whether the graviton, a spin-$2$ particle
that mediates gravity, can have a nonvanishing mass or not. This
question is relevant not only from a theoretical but also from a phenomenological
viewpoint, since a nonzero graviton mass may lead to late-time
acceleration of the universe and thus may be considered as an
alternative to dark energy.

Recently Refs.\cite{deRham:2010ik,deRham:2010kj} proposed the first
example of a fully nonlinear massive gravity theory, where the so called
Boulware-Deser (BD) ghost~\cite{Boulware:1973my}, which had been one of
the major obstacles against a stable nonlinear gravity theory with a
nonvanishing graviton mass, is removed by construction. Due to the
theoretical and phenomenological motivations mentioned above, this
theory has been attracting significant interest.

The first homogeneous and isotropic 
Friedmann-Lema\^\i tre-Robertson-Walker (FLRW) solution for this theory
was presented in \cite{Gumrukcuoglu:2011ew} for Minkowski fiducial
metric and then extended to more general fiducial metrics in
\cite{Gumrukcuoglu:2011zh}. The analysis of linear perturbations in this
general setup was carried out also in
\cite{Gumrukcuoglu:2011zh}. Although a massive spin-$2$ particle  
generically has $5$ propagating degrees of freedom, it was found that
the number of degrees of freedom in the gravity sector was $2$, same as
in general relativity (GR). This is due to the vanishing of the kinetic
terms for the expected additional degrees~\footnote{Backgrounds with
additional symmetries which remove the extra degrees were introduced in
\cite{Tasinato:2012mf}.}. This feature may extend to other setups:
Ref.~\cite{Gratia:2012wt} obtained a vanishing kinetic term on
spherically symmetric inhomogeneous backgrounds.

The goal of the present paper is to determine the fate of the extra
degrees of freedom. We find that in the nonlinear massive gravity, all
cosmological solutions that respect homogeneity and isotropy have a
ghost, i.e.\ an excitation with a wrong sign kinetic term. Therefore,
the universe in this theory should have either
inhomogeneities~\cite{D'Amico:2011jj} or anisotropy~\cite{GLM2012}.  We
note that the ghost mode found in the present paper is among five
degrees of freedom of the massive spin-$2$ field and thus, is not the BD
ghost.

{\bf The model and the background.~}
Imposing the absence of the BD ghost, the massive gravity action, in
vacuum, can be constructed as \cite{deRham:2010kj} 
\begin{equation}
S = \frac{M_{\mathrm{Pl}}^2}2\int d^4 x\,\sqrt{-g}[R-2\Lambda+2m_g^2\mathcal{L}_{\rm MG}]\,,
\end{equation}
with $\mathcal{L}_{\rm MG}={\cal L}_2 + \alpha_3\,{\cal L}_3 + \alpha_4 \, {\cal L}_4$, where
\begin{eqnarray*}
 {\cal L}_2 & = & \tfrac{1}{2}
  ([{\cal K}]^2-[{\cal K}^2]),\\
 {\cal L}_3 & = & \tfrac{1}{6}
  ([{\cal K}]^3-3[{\cal K}][{\cal K}^2]+2[{\cal K}^3]),\\
 {\cal L}_4 & = & \tfrac{1}{24}
  ([{\cal K}]^4-6[{\cal K}]^2[{\cal K}^2]+3[{\cal K}^2]^2+8[{\cal K}][{\cal K}^3]-6[{\cal K}^4]),
\label{lag234}
\end{eqnarray*}
the square brackets denote the trace operation, and
\begin{equation}
{\cal K}^\mu{}_\nu = \delta^\mu{}_\nu 
 - \bigl(\sqrt{g^{-1}f}\bigr)^{\mu}{}_{\nu}\,.
\label{Kdef}
\end{equation}
Here, $g_{\mu\nu}$ and $f_{\mu\nu}$ are physical and
fiducial metrics, respectively. Since we are interested in the stability of the gravity sector only, it is sufficient to consider a vacuum configuration, with a cosmological constant $\Lambda$.

The physical metric is chosen to be the simplest anisotropic extension
of FLRW, namely, the axisymmetric Bianchi type--I metric 
\begin{equation}
ds^2 = - N^2 dt^2 + a^2(e^{4\,\sigma}\,dx^2 + e^{-2\,\sigma}\,\delta_{ij}\,dy^i\,dy^j)\,,
\label{eq:g0mn}
\end{equation}
where $N$, $a$, and $\sigma$ are functions of the time variable $t$. In
the rest of the paper, Greek indices span the space-time coordinates,
while the indices $i,j=2,3$ correspond to the coordinates on the
$y$--$z$ plane, with $y^2=y$, $y^3 = z$. Since our goal is to obtain the
stability conditions of this metric in the isotropic limit,
the whole system in this limit needs to reduce to the general
cosmological solutions given in \cite{Gumrukcuoglu:2011ew,
Gumrukcuoglu:2011zh}. For this reason, we consider a fiducial metric to
be in the flat FLRW form, 
\begin{equation}
f_{\mu\nu} = -n^2 \partial_\mu \phi^0 \partial_\nu \phi^0+ \alpha^2 (\partial_\mu\phi^1\partial_\nu \phi^1+\delta_{ij}\partial_\mu\phi^i\partial_\nu \phi^j),
\label{eq:fmn}
\end{equation}
where both $n$ and $\alpha$ are functions of the time-St\"uckelberg
field $\phi^0$. 

The equations of motion for the background can be calculated by varying the action with respect to the St\"uckelberg fields and the metric. As a result, we obtain three independent equations as
\begin{widetext}
\begin{eqnarray}
&&3\left(H^2 - \Sigma^2\right)-\Lambda = m_g^2\left[ -(3\,\gamma_1-3\,\gamma_2+\gamma_3)+
 \gamma_1\,(2\,e^\sigma +e^{-2\sigma})X- \gamma_2(e^{2\sigma}+2\,e^{-\sigma})\,X^2 +\gamma_3\,X^3\right],\label{eq:fried}\nonumber\\
&&\frac{3\dot{\Sigma}}{N}+9H\Sigma =
m_g^2 (e^{-2\,\sigma}-e^\sigma)X\left[ \gamma_1 -\gamma_2(e^\sigma+r)X+ \gamma_3\,r e^\sigma X^2\right],\nonumber\\
&&J_\phi^{(x)}\,(H+2\Sigma -H_f\,e^{-2\sigma}X)+2J_\phi^{(y)}\,(H-\Sigma-H_fe^{\sigma}X)=0\,,
\label{eq:stuckfieldeq}
\end{eqnarray}
\end{widetext}
where
\begin{eqnarray}
J_\phi^{(x)}  &\equiv& \gamma_1-2\,\gamma_2\,e^\sigma\,X+\gamma_3\,e^{2\sigma}\,X^2\,,\nonumber\\
J_\phi^{(y)}  &\equiv& \gamma_1-\,\gamma_2\,(e^{-2\sigma}+e^\sigma)\,X+\gamma_3\,e^{-\sigma}\,X^2\,,
\label{eq:jabdef}
\end{eqnarray}
and 
\begin{align}
&\gamma_1\equiv3+3\,\alpha_3+\alpha_4\,,\;
\gamma_2\equiv1+2\,\alpha_3+\alpha_4\,,\;
\gamma_3\equiv\alpha_3+\alpha_4
\nonumber\\
&\qquad \quad H\equiv {\dot{a}}/({aN})\,,\;
H_f\equiv{\dot{\alpha}}/({\alpha n})\,,\;
\Sigma \equiv \dot{\sigma}/{N}\,,\nonumber\\
&\qquad\qquad\qquad  X \equiv{\alpha}/{a}\,,\;
r\equiv an/(\alpha N).
\end{align}

We note that, in the isotropic limit $\left(\sigma,\,\Sigma\to0\right)$,
we have $J_\phi^{(x)}=J_\phi^{(y)}$, so that the St\"uckelberg equation
of motion, Eq.~(\ref{eq:stuckfieldeq}), at leading order, gives 
\begin{equation}
\gamma_1-2\,\gamma_2\,X + \gamma_3\,X^2\simeq0\,,
\end{equation}
that is $X\to{\rm constant}$, which corresponds to the FLRW result found
in \cite{Gumrukcuoglu:2011zh}. In the same limit, we can also see that
$H\to{\rm constant}$, as expected. 

{\bf Even modes.~}
Let us now consider the perturbations which transform as 2D scalars under a
spatial rotation in the $y{-}z$ plane (also referred as even
modes). Then, the perturbed metric for the even sector can be written as \cite{Gumrukcuoglu:2007bx}
\begin{eqnarray}
ds^2 &=& -N^2(1+2\Phi)dt^2 + 2aNdt[e^{2\sigma}\partial_x \chi dx+ e^{-\sigma}\partial_i B dy^i]
\nonumber\\
&&{}+ a^2e^{4\sigma}(1+\psi)dx^2 
+2a^2e^\sigma\partial_x\partial_i\beta dxdy^i \nonumber\\
&&{}+ a^2e^{-2\sigma}[\delta_{ij}(1+\tau) +\partial_i\partial_j E]dy^idy^j\,,
\end{eqnarray}
while the even-type perturbations of St\"uckelberg fields read
\begin{equation}
\phi^0 = t + \pi^0\,,\;
\phi^1 = x + \partial_x\pi^1\,,\;
\phi^i = y^i + \partial^i \pi\,.
\end{equation}
We can then define gauge invariant combinations as follows
\begin{eqnarray}
\hat{\Phi} &=& \Phi- \frac{1}{2\,N}\,\partial_t\left(\frac{\tau}{H-\Sigma}\right)\,,\nonumber\\
\hat{\chi} &=& \chi + \frac{\tau\, e^{-2\sigma}}{2a(H-\Sigma)}- \frac{ae^{2\sigma}}{N}\partial_t\!\left[ e^{-3\sigma}\!\left(\beta - \frac{e^{-3\sigma}}{2}E\right)\right],\nonumber\\
\hat{B} &=& B + \frac{e^\sigma}{2\,a\,(H-\Sigma)}\,\tau - \frac{a\,e^{-\sigma}}{2\,N}\,\dot{E}\,,\nonumber\\
\hat{\psi} &=& \psi -\frac{H+2\,\Sigma}{H-\Sigma}\,\tau - e^{-3\,\sigma}\,\partial_x^2\left(2\,\beta - e^{-3\,\sigma}\,E\right)\,,\nonumber\\
\hat{\tau}_\pi &=& \pi^0 - \frac{\tau}{2\,N\,(H-\Sigma)}\,,\nonumber\\
\hat{\beta}_\pi &=& \pi^1 - e^{-3\,\sigma}\,\left(\beta - \frac{e^{-3\,\sigma}}{2}\,E\right)\,,\nonumber\\
\hat{E}_\pi &=& \pi -\frac{1}{2}\,E\,. 
\label{eq:gi-gr-even}
\end{eqnarray}
The first four definitions do not refer to the St\"uckelberg perturbations and are thus already present in GR \cite{Himmetoglu:2008zp}. However, the additional three degrees arise from the breaking of general coordinate invariance by the non zero expectation value of the St\"uckelberg fields.

In order to find the behavior of the perturbations, we proceed as usual
by expanding the action at second order in the perturbation fields, then by employing the 
Fourier plane-wave decompositions, as in 
$\exp[i(k_L x+k_i y^i)]$. The degrees of freedom arising from the $g_{0\mu}$
perturbations, namely $\hat{\Phi}$, $\hat{B}$ and $\hat{\chi}$, are
nondynamical, thus can be integrated out. Furthermore, the kinetic term
for the $\hat{\tau}_\pi$ is proportional to the background equations of
motion, so that this degree of freedom is also nondynamical. We
interpret this field as the would-be BD ghost, which is eliminated in
this theory by construction.

In the massless theory (i.e.\
GR), using the constraint equations also removes the
degrees $\hat{\beta}_\pi$, $\hat{E}_\pi$, leaving only $\hat{\psi}$ in
the action, which becomes one of two gravity wave polarizations in the
isotropic limit. However, in our case, due to the nonzero mass of the
graviton, these two degrees of freedom are dynamical, in
general.

Thus, the Lagrangian for even-type perturbations in vacuum
has three physical propagating modes, ${\cal V}_a$, ($a=1,2,3$). Assuming small deviation from FLRW, with $|\sigma|\ll1$ and $|\Sigma/H|\ll 1$, we study the kinetic matrix ${\cal K}_{ab}$
\begin{equation}
S^{(2)}_{\rm even}\ni \frac{M_p^2}{2}\,\int N\,dt\,dk_L\,d^2k_T\,a^3 \left(\frac{\dot{\cal V}_a^\star}{N} {\cal K}_{ab} \frac{\dot{\cal V}_b}{N}\right)\,.
\end{equation}
Thanks to the 2D rotational symmetry on the $y$--$z$ plane, the action depends on $k_T\equiv\sqrt{k_2^2+k_3^2}$, instead of the individual components. The eigenvalues of ${\cal K}_{ab}$, at leading order in small anisotropy expansion, are
\begin{equation}
\kappa_1 \simeq \frac{p_T^4}{8\,p^4}\,,\;
\kappa_2 \simeq - \frac{2a^4M_{\rm GW}^2p_L^2}{1-r^2}\sigma\,,\;
\kappa_3 \simeq -\frac{p_T^2}{2\,p_L^2}\,\kappa_2\,,
\label{kineig}
\end{equation}
where we defined $M_{\rm GW}^2 \equiv m_g^2(1-r)X^2(\gamma_2-\gamma_3X)$, and introduced the physical momenta 
\begin{equation}
p_L \equiv \tfrac{k_L}{ae^{2\sigma}}\simeq \tfrac{k_L}{a}\,,\;
p_T \equiv \tfrac{k_T}{ae^{-\sigma}}\simeq \tfrac{k_T}{a}\,,\; p^2 \equiv p_L^2 + p_T^2\,.
\end{equation}
The kinetic term $\kappa_1$ which is the only eigenvalue that does not vanish in isotropic limit, corresponds to one of the gravity wave polarizations in FLRW. Once small but nonvanishing anisotropy is introduced, two additional even modes acquire nonzero kinetic terms at quadratic order. More importantly, from (\ref{kineig}), we see that $\kappa_2$ and $\kappa_3$ have opposite signs, regardless of the parameters of the theory. Thus, we conclude that in the isotropic limit, one of the new degrees is always a ghost. 
Assuming that $\sigma(1-r)>0$ (which turns out to be the condition for stability in the odd sector, as we show later), the ghost mode is associated with the eigenvalue $\kappa_2<0$. 

We conclude the discussion of the even modes by presenting their dispersion
relations. We first make a field redefinition into new field basis
fields ${\cal W}_a $ defined such that the kinetic action can be
written as  
\begin{equation}
S^{(2)}_{\rm even} \ni \frac12\int N\,dt \,dk_L d^2k_T a^3 \left(\frac{\dot{\cal W}_a^\star}{N} \eta_{ab} \frac{\dot{\cal W}_b}{N}\right)\,, \end{equation}
 where $\eta_{ab} = {\rm diag}(1,-1,1)$. The mass spectrum can be determined either by 
studying the equation for the frequency-discriminant, or equivalently,
by performing a Lorentz transformation to diagonalize the frequency
matrix. Eventually, we find 
\begin{eqnarray}
\omega_1^2 &\simeq& p^2+ M_{\rm GW}^2 \,,\nonumber\\
\omega_2^2 &\simeq& -\frac{1-r^2}{24\sigma}\!\!\left[\sqrt{(10p^2+p_T^2)^2 - 8p_L^2p_T^2} - (2p^2 +3p_T^2) \right],\nonumber\\
\omega_3^2 &\simeq& -\omega_2^2+\frac{1-r^2}{12\sigma}\,(2p^2+3p_T^2)\,,
\end{eqnarray}
with $\omega_2^2\omega_3^2<0$ in general, and $\omega_2^2<0$ by assuming $\sigma(1-r)>0$. We note that the dispersion relation corresponding to the ghost,  $\omega_2^2$, becomes smaller at larger scales. Therefore, at sufficiently large scales, this mode cannot
be integrated out from the low energy effective theory. This feature makes the FLRW background unstable for massive gravity. As a
consequence, the homogeneous and isotropic cosmology cannot be accommodated in the nonlinear massive gravity theory.

{\bf Odd modes.~} Let us now discuss the odd sector (i.e.\ the divergence-less part of the
modes which transform as 2D vectors under a rotation in the $y{-}z$
plane). The perturbed metric we consider is 
\begin{eqnarray}
ds^2 &=&-N^2dt^2 +2ae^{-\sigma}Nv_idtdy^i+2a^2e^\sigma\partial_x\lambda_i dxdy^i\nonumber\\
&&{}+a^2e^{4\sigma} dx^2+a^2e^{-2\sigma}(\delta_{ij}+\partial_{(i}h_{j)})dy^idy^j\,,
\label{eq:oddmetric}
\end{eqnarray}
where $\partial_{(i}h_{j)} \equiv (\partial_i h_j + \partial_i h_j)/2$
and $\partial^i v_i = \partial^i \lambda_i = \partial^i h_i =0$. For the 
St\"uckelberg fields, we consider instead 
\begin{equation}
\phi^0=t\,,\,\phi^1=x\,,\,\phi^i=y^i + \pi^i\,,
\label{eq:oddstuck}
\end{equation}
where  $\partial_i \pi^i=0$. 
Since the vectors are defined on the 2D $y$--$z$ plane, the transverse
condition can be used to reduce each of these vectors to a single degree
of freedom 
\begin{equation*}
v_i = \epsilon_i^{\;\,j}\partial_jv ,\;
\lambda_i = \epsilon_i^{\;\,j}\partial_j\lambda,\;
h_i = \epsilon_i^{\;\,j}\partial_jh ,\;
\pi_i = \epsilon_i^{\;\,j}\partial_j\pi_{\rm odd},
\end{equation*}
where $\epsilon_i^{\;\,j}$ is a unit anti-symmetric tensor with $\epsilon_2^{\;\,3}= -\epsilon_3^{\;\,2}= 1$.
Also for the odd modes we can introduce gauge invariant combinations as follows
\begin{eqnarray}
\hat{v} &=& v - \frac{a\,e^{-2\,\sigma}}{2\,N}\,\dot{h}\,,\nonumber\\
\hat{\lambda} &=& \lambda - \frac{e^{-3\,\sigma}}{2}\,h\,,
\label{eq:gi-gr-odd}\nonumber\\
\hat{h}_\pi &=& \pi_{\rm odd} - \frac{1}{2}\,h\,.
\label{eq:gi-st-odd}
\end{eqnarray}
Using these fields, the second-order resulting action depends on the
three perturbations ($\hat{v}$, $\hat{\lambda}$, $\hat{h}_\pi$). Among these,
$\hat{v}$ does not have any time derivatives and can be removed by
solving its own constraint equation. In General Relativity, this
operation also removes $\hat{h}_\pi$ and the final action can be written in
terms of $\hat{\lambda}$ only. However, in this nonlinear theory of
massive gravity, we expect the field $\hat{h}_\pi$ to remain in the action as
an extra degree of freedom coming from the St\"uckelberg sector. 

After a further field redefinition, 
\begin{equation}
{\cal Q}_1 \equiv -e^{3\,\sigma}\,\hat{\lambda} \,,
\qquad
{\cal Q}_2 \equiv \frac{2\,e^{3\,\sigma}\,p_L^2}{p^2}\,\hat{\lambda} - 2\,\hat{h}_\pi\,,
\end{equation}
the quadratic action, for small anisotropy, takes the following form 
\begin{eqnarray}
S^{(2)}_{\rm odd} &\simeq& \frac{M_{\rm Pl}^2}{2}\int N\,dt\,dk_Ld^2 k_Ta^3
\left[K_{11} \frac{\vert\dot{\cal Q}_1\vert^2}{N^2}
- \Omega^2_{11}\,\left\vert{\cal Q}_1\right\vert^2\right.\nonumber\\
&&\qquad\qquad\qquad\quad{}+\left.K_{22}\, \frac{\vert\dot{\cal Q}_2\vert^2}{N^2}
- \Omega^2_{22}\,\left\vert{\cal Q}_2\right\vert^2\right],
\end{eqnarray}
where the two modes decouple at leading order in the small anisotropy expansion, with coefficients 
\begin{eqnarray}
K_{11} &=& \frac{a^4\,p_L^2\,p_T^4}{2\,p^2}\,,\quad
K_{22} = \frac{a^4\,p_T^2\,M_{GW}^2}{4\,(1-r^2)}\,\sigma\,,\nonumber\\
\frac{\Omega^2_{11}}{K_{11}} &=& p^2+M^2_{\rm GW}\,,\quad\frac{\Omega^2_{22}}{K_{22}} = c_{\rm odd}^2\,p^2\,,
\end{eqnarray}
and $c_{\rm odd}^2 = (1-r^2)/(2\sigma)$. Thus, at leading order, we
identify the mode ${\cal Q}_1$ with one of the gravity wave
polarizations in the FLRW background \cite{Gumrukcuoglu:2011zh}. The
extra degree of freedom ${\cal Q}_2$ is massless and has sound speed
$c_{\rm odd}$. In order for this mode to be stable, we
require the kinetic term  for ${\cal Q}_2$ to be positive, that is 
\begin{equation}
(1-r)\,\sigma >0\,.
\label{eq:noghost}
\end{equation}
In this case, also $c_{\rm odd}^2$ becomes positive, and the odd mode
${\cal Q}_2$ is, in general, free from ghost instabilities.

{\bf Conclusions.~} 
In the search for a theory which could explain the dark energy enigma,
the nonlinear massive gravity, recently introduced in
\cite{deRham:2010ik}, has raised lots of interest among both theoretical
and experimental physicists, 
thanks to its implications in our
understanding of fundamental forces, if it is theoretically consistent and
observationally viable. 

This theory admits homogeneous and isotropic solutions, and it has been
shown in \cite{Gumrukcuoglu:2011zh} that, out of the five modes which
would be typically expected in this theory, only two actually
propagate at the linearized level. Therefore, it is of interest to investigate
the reason for this unexpected feature. 

We propose here that this phenomenon is due to the high symmetry
structure of the FLRW background. Accordingly, we have studied the small 
anisotropy limit of the Bianchi--I manifold and found that 
there is always a ghost mode in the even sector, with a propagation speed 
that diverges in the isotropic limit. Furthermore, this mode does 
not have a mass gap; its frequency tends to zero for small values of the 
momentum. Therefore, at sufficiently large scales, the frequency cannot be 
considered as large compared to the ultraviolet cutoff of the theory. As
a consequence, the ghost mode cannot be integrated out in general, and the 
almost-isotropic background becomes unstable under production of
negative energy quanta.

Although our analysis is linear, the terms in the quadratic action with
coefficients proportional to the small anisotropy can be interpreted as
the leading order nonlinear perturbations in a pure FLRW universe. The
presence of a mode with negative kinetic term indicates that a
homogeneous and isotropic universe in nonlinear massive gravity is
unstable.

Our conclusion about ghost instability is far more general than it 
appears~\footnote{Partially massless gravity \cite{deRham:2012kf} 
may evade our conclusion but it is a different theory. Also, nonlinear
completion is not known.}, despite the simplicity of the analysis presented
above. This is because, whenever a quadratic kinetic term vanishes, the
leading kinetic term is generically cubic and thus can easily become
negative, signaling the existence of ghost at the nonlinear
level. Moreover, the other type of homogeneous and isotropic solutions
(in the non-self-accelerating branch) suffer from ghost instability
already at the linearized level \cite{Fasiello:2012rw}, as expected from
the classical work of Higuchi \cite{Higuchi:1986py}. Therefore, all
homogeneous and isotropic backgrounds, as well as most (if not all) of
known spherically-symmetric inhomogeneous solutions, are unstable.

~\\~
\begin{acknowledgments}
 The work of A.E.G.\ and S.M.\ was supported by the World Premier
 International Research Center Initiative (WPI Initiative), MEXT,
 Japan. S.M.\ also acknowledges the support by Grant-in-Aid
 for Scientific Research 24540256 and 21111006, and by Japan-Russia
 Research Cooperative Program. The authors are grateful to Yukawa
 Institute for warm hospitality during the workshop YITP-W-11-26, when
 this project was initiated. 
\end{acknowledgments}


\end{document}